\documentclass[twocolumn]{aastex61}
\pdfoutput=1 
\usepackage{amsmath,amstext}
\usepackage[T1]{fontenc}
\usepackage{apjfonts} 
\usepackage[figure,figure*]{hypcap}

\newcommand {\nhunit} {cm$^{-2}$}
\newcommand {\nh} {$N_{\mathrm{H}}$}

\shorttitle{Local AGN Clustering}
\shortauthors{Powell M et al.}

\begin{document}

\title{The Swift/BAT AGN Spectroscopic Survey -- IX. The clustering environments of an unbiased sample of Local AGN}

\author{M. C. Powell}
\affiliation{Yale Center for Astronomy and Astrophysics, and Physics Department, Yale University, P.O. Box 2018120, New Haven, CT 06520-8120, USA}

\author{N. Cappelluti}
\affiliation{Yale Center for Astronomy and Astrophysics, and Physics Department, Yale University, P.O. Box 2018120, New Haven, CT 06520-8120}
\affiliation{Physics Department, University of Miami, Coral Gables, FL 33155, USA}

\author{C.M. Urry}
\affiliation{Yale Center for Astronomy and Astrophysics, and Physics Department, Yale University, P.O. Box 2018120, New Haven, CT 06520-8120}

\author{M. Koss}
\affiliation{Eureka Scientific, 2452 Delmer Street, Suite 100, Oakland, CA 94602-3017, USA}

\author{A. Finoguenov}
\affiliation{Department of Physics, University of Helsinki, FI-00014 Helsinki, Finland}
\affiliation{Max-Planck-Institut f{\"u}r extraterrestrische Physik, Giessenbachstrasse 1, D-85748 Garching, Germany}

\author{C. Ricci}
\affiliation{N\'ucleo de Astronom\'ia de la Facultad de Ingenier\'ia, Universidad Diego Portales, Av. Ej\'ercito Libertador 441, Santiago, Chile.}
\affiliation{Instituto de Astrof\'{\i}sica, Facultad de F\'{i}sica, Pontificia Universidad Cat\'{o}lica de Chile, Casilla 306, Santiago 22, Chile.}
\affiliation{Chinese Academy of Sciences South America Center for Astronomy and China-Chile Joint Center for Astronomy, Camino El Observatorio 1515, Las Condes, Santiago, Chile}
\affiliation{Kavli Institute for Astronomy and Astrophysics, Peking University, Beijing 100871, China}

\author{B. Trakhtenbrot}
\affiliation{Department of Physics, ETH Zurich, Wolfgang-Pauli-Strasse 27, CH-8093 Zurich, Switzerland}
\affiliation{Zwicky Fellow}

\author{V. Allevato}
\affiliation{Scuola Normale Superiore, Piazza dei Cavalieri 7, I-56126 Pisa, Italy}
\affiliation{Helsinki Institute of Physics, PO Box 64, 00014 Helsinki, Finland}

\author{M. Ajello}
\affiliation{Department of Physics and Astronomy, Clemson University, Clemson, SC 29634-0978, USA}

\author{K. Oh}
\affiliation{Institute for Particle Physics and Astrophysics, Department of Physics, ETH Zurich, Wolfgang-Pauli-Strasse 27, CH-8093 Zurich, Switzerland}
\affiliation{Department of Astronomy, Kyoto University, Oiwake-cho, Sakyo-ku, Kyoto 606-8502, Japan}
\affiliation{JSPS Fellow}

\author{K. Schawinski}
\affiliation{Institute for Particle Physics and Astrophysics, Department of Physics, ETH Zurich, Wolfgang-Pauli-Strasse 27, CH-8093 Zurich, Switzerland}

\author{N. Secrest}
\affiliation{U.S. Naval Observatory, 3450 Massachusetts Avenue NW, Washington, DC 20392, USA}

\begin{abstract}

We characterize the environments of local accreting supermassive black holes by measuring the clustering of AGN in the {\it Swift}/BAT Spectroscopic Survey (BASS). With 548 AGN in the redshift range $0.01<z<0.1$ over the full sky from the DR1 catalog, BASS provides the largest, least biased sample of local AGN to date due to its hard X-ray selection (14-195 keV) and rich multiwavelength/ancillary data. By measuring the projected cross-correlation function between the AGN and 2MASS galaxies, and interpreting it via halo occupation distribution and subhalo-based models, we constrain the occupation statistics of the full sample, as well as in bins of absorbing column density and black hole mass. We find that AGN tend to reside in galaxy group environments, in agreement with previous studies of AGN throughout a large range of luminosity and redshift, and that on average they occupy their dark matter halos similar to inactive galaxies of comparable stellar mass.
We also find evidence that obscured AGN tend to reside in denser environments than unobscured AGN, even when samples were matched in luminosity, redshift, stellar mass, and Eddington ratio. We show that this can be explained either by significantly different halo occupation distributions or statistically different host halo assembly histories. 
Lastly, we see that massive black holes are slightly more likely to reside in central galaxies than black holes of smaller mass. 
\end{abstract}

\section{Introduction}
Studying the large-scale environments of Active Galactic Nuclei (AGNs) is important for understanding the growth of supermassive black holes (SMBHs) and how they coevolve with their host galaxies \citep[e.g.,][]{Kormendy:2013}. Clustering is a powerful tool in statistically determining the typical dark matter halo in which AGN reside, as well as how they occupy their halos. Coupled with a sensible model of halo mass assembly, this can constrain fueling mechanisms (i.e., mergers versus secular evolution) and feedback scenarios, providing selection effects are properly taken into account.

Previous studies of AGN clustering using soft X-ray and optically selected samples have found somewhat discrepant results for the typical host halo mass of AGN. Luminous quasars drawn from wide-area optical surveys appear to lie in smaller halos ($M_{\rm{h}}\sim 10^{12.5}$ $M_{\odot}h^{-1}$, where $h=H_0/100$ km s$^{-1}$ Mpc$^{-1}$) than moderate-luminosity X-ray AGN found in deeper surveys ($M_{\rm{h}}\sim 10^{13}$ $M_{\odot}h^{-1}$)  across a wide range of redshift \citep[e.g.,][]{Croom:2005,Gilli:2005,Gilli:2009,Krumpe:2012,Ross:2009,Shen:2009,Allevato:2011,Allevato:2014}. Additionally, it is not clear whether unobscured and obscured AGN (either defined by their column density or optical classification) have the same clustering statistics, in accordance with the unified model, or if they tend to reside in different environments due to different accretion modes or due to being two stages of one evolutionary track, as claimed by recent (but discordant) studies \citep{Allevato:2014,Villarroel:2014,Mendez:2016,DiPompeo:2017}. However, they all probe different volumes, host galaxy properties, and luminosity ranges, making comparison between studies difficult (see, e.g., \citealt{Mendez:2016}). Additionally, the picture may be confused because a large number of obscured AGN have been missed in optical and soft X-ray surveys due to dust and gas obscuration from the torus and/or host galaxy. Population synthesis models of the Cosmic X-ray background indicate that a significant fraction of SMBH accretion occurs in obscured environments \citep{Treister:2004,Treister:2012}, meaning obscured AGN are a vital population to consider in a full model of halo, galaxy, and SMBH (co-)evolution. Hard X-ray selection ($>10$ keV) can remedy this obscuration-related bias, as the majority of energetic photons are able to pass through large columns of gas and dust, up to Compton-thick levels (\nh$\approx10^{24}$\nhunit; \citealt{Ricci:2015}). In addition, hard X-ray selection is extremely efficient, as there are very few contaminates, including the host galaxy.

The Burst Alert Telescope (BAT; \citealt{Barthelmy:2005,Krimm:2013}) instrument on the {\it Swift} satellite \citep{Gehrels:2004} has surveyed the entire sky to unprecedented sensitivity in the 14-195 keV band \citep{Baumgartner:2013,Oh:2018}. Local AGN detected by BAT include the obscured and/or low-luminosity AGN missed by optical detection, as well as the rare high-luminosity AGN only found in wide-area surveys, so that BAT AGN can solve some of the aforementioned issues with previous AGN clustering studies. \cite{Cappelluti:2010} were the first to measure the clustering of \emph{Swift}/BAT AGN using a sample of 199 AGN in the 36-month catalog \citep{Ajello:2009}, but had uncertain results due to the small sample size. While they did find a dependence in X-ray luminosity, it was most likely a selection effect due to the strong redshift dependence inherent in any small flux-limited sample.

In this study, we have more than doubled the sample by using the 70-month {\it Swift}/BAT AGN catalog \citep{Baumgartner:2013}, 
along with spectroscopic information from the {\it Swift}/BAT Spectroscopic Survey (BASS; \citealp{Koss:2017}),
to constrain the AGN halo occupation distribution (HOD) for 499 BASS AGN in the redshift range $0.01<z<0.1$. To improve the statistics, we cross-correlate the AGN with local 2MASS galaxies that trace the underlying dark matter distribution. Additionally, we investigate the environmental dependence of AGN parameters like obscuring column density and black hole mass, while matching distributions in X-ray luminosity, redshift, stellar mass, and Eddington ratio.

\cite{Krumpe:2017} recently published a similar, independent clustering analysis of \emph{Swift}/BAT AGN, in which they analytically fit the cross-correlation function with 2MASS galaxies. They also divided their sample by optical classification (Type 1 or Type 2) from \cite{Baumgartner:2013}, as well as by observed X-ray luminosity. 
However, detailed X-ray spectral fitting \citep{Ricci:2017B} allows us to estimate the \emph{intrinsic} absorption-corrected luminosity for each AGN, which differs strongly from the observed value at $N_{\rm H}>10^{23.5}$ cm$^{-2}$.
The BASS DR1 release \citep{Koss:2017} also includes 46 new redshifts for a spectroscopic completeness of over 95\%, and provides column densities for each 836 AGN, which is the method for understanding whether an AGN is obscured or not.
Our study also differs from \cite{Krumpe:2017} in how we fit models; namely, we populate dark matter halos statistically from $N$-body simulations (using the \texttt{Halotools} software package; \citealt{Hearin:2017}). Because this allows a straightforward correction for catalog incompleteness, we use an extended redshift range ($z<0.1$ rather than $ z<0.037$) for better number statistics (499 AGN compared to 274 in \citeauthor{Krumpe:2017}), and we do not have to rely on assumptions from analytic models. The simulation-based approach also allows us to look beyond halo mass to other halo parameters like halo concentration, in order to investigate effects such as assembly bias. 
In this paper we challenge the idea that AGN clustering is driven only by the typical mass of its dark matter halo.

This paper is organized as follows: we describe the data selection of the BASS AGN and 2MASS galaxies in Section 2; our method for measuring the correlation function and fitting it with a halo model is described in Section 3; Section 4 presents the results for the full AGN sample, as well as the dependence on obscuration and black hole mass; we discuss our findings in Section 5, and summarize them in Section 6.  
We assume flat $\Lambda$CDM cosmology ($\Omega_{m}=0.3$, $\Omega_{\Lambda}=0.7$, $H_{0}=100~h^{-1}$ km s$^{-1}$ Mpc$^{-1}$, $h=0.7$), and errors quoted are $1 \sigma$ unless otherwise stated.
\section{Data}
\subsection{AGN Sample}

BASS consists of 836 local AGN from the $Swift$-BAT 70-month catalog \citep{Koss:2017,Ricci:2017B}, selected by their hard X-ray emission ($14-195$ keV), which has the benefit of being unbiased toward obscuration. BASS comprises the largest, most unbiased sample of local AGN to date, and there is an abundance of complementary multiwavelength ancillary data\footnote{www.bass-survey.com}. 

Each AGN has soft X-ray data from {\it Chandra}, {\it XMM-Newton}, {\it Suzaku}, or {\it Swift}/XRT, so that the full X-ray spectra have been modeled ($0.3-150$ keV; \citealt{Ricci:2017B}). These give the obscuring column ($N_{\rm H}$) and intrinsic X-ray flux for each AGN, from which bolometric luminosities have been estimated using a fixed hard X-rays bolometric correction to the $14-195$ keV luminosities ($L_{\rm{bol}}=8~L_{14-195~keV}$; \citealt{Koss:2017}).

Optical spectroscopy has been obtained for 641 unbeamed AGN, providing spectroscopic redshifts that allow for 3D clustering analyses. We assume that the AGN without spectra ($5\%$) do not systematically affect the clustering of the overall population, as we verified negligible differences between the flux distributions and angular correlation functions with and without their inclusion. Black hole masses have been estimated for 429 AGN, of which 54\% are unobscured and 46\% are obscured. Black hole masses from unobscured AGN were estimated from H$\beta$ and/or H$\alpha$ broad lines FWHM \citep{Kaspi:2000,Greene:2005,Bentz:2009,Trakhtenbrot:2012,MR:2016}; these have uncertainties of $0.3-0.4$ dex \cite[e.g.,][]{Shen:2013,Peterson:2014}. For obscured AGN without broad lines, we relied on the $M_{\rm BH}-\sigma_{*}$ relation \citep{Kormendy:2013}, where $\sigma_{*}$ was measured by fitting the spectra with host galaxy stellar templates. These black hole mass estimates have slightly larger uncertainties of $\sim0.5$ dex \citep{Xiao:2011}. Eddington ratios ($\lambda_{\rm{Edd}}\equiv L_{\rm{bol}}/L_{\rm{Edd}}$) were derived from the bolometric luminosities and black hole masses via $L_{\rm{bol}}/1.3\times 10^{38}$erg s$^{-1}[M_{\rm{BH}}/M_{\odot}]$.
The uncertainties on $\lambda_{\rm{Edd}}$ are driven by the large systematic uncertainties on both $M_{\rm{BH}}$ determinations (up to $\sim$0.5 dex, see above) and bolometric corrections. The latter may be of roughly 0.2-0.3 dex, and perhaps involve more complicated uncertainties related to possible trends with luminosity and/or $\lambda_{\rm{Edd}}$ itself \citep{Marconi:2004,Vasudevan:2007,Jin:2012}.
 More details of the optical spectral analysis can be found in \cite{Koss:2017}.

Stellar masses of the BAT AGN host galaxies were derived by spectrally de-convolving the AGN emission from stellar emission via SED decomposition. We combined near-IR data from 2MASS, which is more sensitive to stellar emission, with mid-IR data from the AllWISE catalog \citep{Wright:2010}, which is more sensitive to AGN emission. Where available, isophotal near-IR magnitudes from the 2MASS XSC were used to capture the most stellar emission, and the corresponding AllWISE elliptical magnitudes were used. We then converted the magnitudes to the AB system, and corrected for Galactic reddening using $E(B-V)$ estimates from \citet{Schlafly:2011}. We used the low-resolution SED templates from \citet{Assef:2010} to decompose the BAT AGN host galaxies into a linear combination of an AGN plus early-type (E), continuously star-forming (Sbc), and starburst galaxies (Im). To convert the luminosities of the galaxy components to masses, we obtained their stellar mass coefficients by fitting them with the \citet{Blanton:2007} stellar population synthesis templates. The templates were convolved with the 2MASS/WISE system responses, and fit to the data via weighted non-negative least-squares, where for the weights we use the inverse variances of the data. Finally, we include AGN reddening by performing the SED decompositions along a logarithmically spaced grid of $E(B-V)$ values, choosing the value that yields the lowest $\chi^2$. To estimate random errors in the stellar mass uncertainties, we re-fit each source several times, each time removing one of the seven photometric data points (jackknife resampling), and permuting the remaining magnitudes by their uncertainties. This produced random errors of about 0.06 dex. There is also a component of scatter introduced by using the \citet{Assef:2010} templates, which have three stellar components, instead of the five original \citet{Blanton:2007} stellar population templates. By fitting the NASA-Sloan Atlas photometry with the \citet{Assef:2010} templates, we estimate that there is an additional scatter term of about 0.08 dex, which we add in quadrature to the random error term provided above. Finally, the absolute stellar mass uncertainty for masses estimated using near-IR photometry is approximately a factor of two \citep{BelldeJong:2001}. We therefore estimate that our stellar mass uncertainties are about 0.32 dex, on average. 

\begin{figure}
\centering
\includegraphics[width=.5\textwidth]{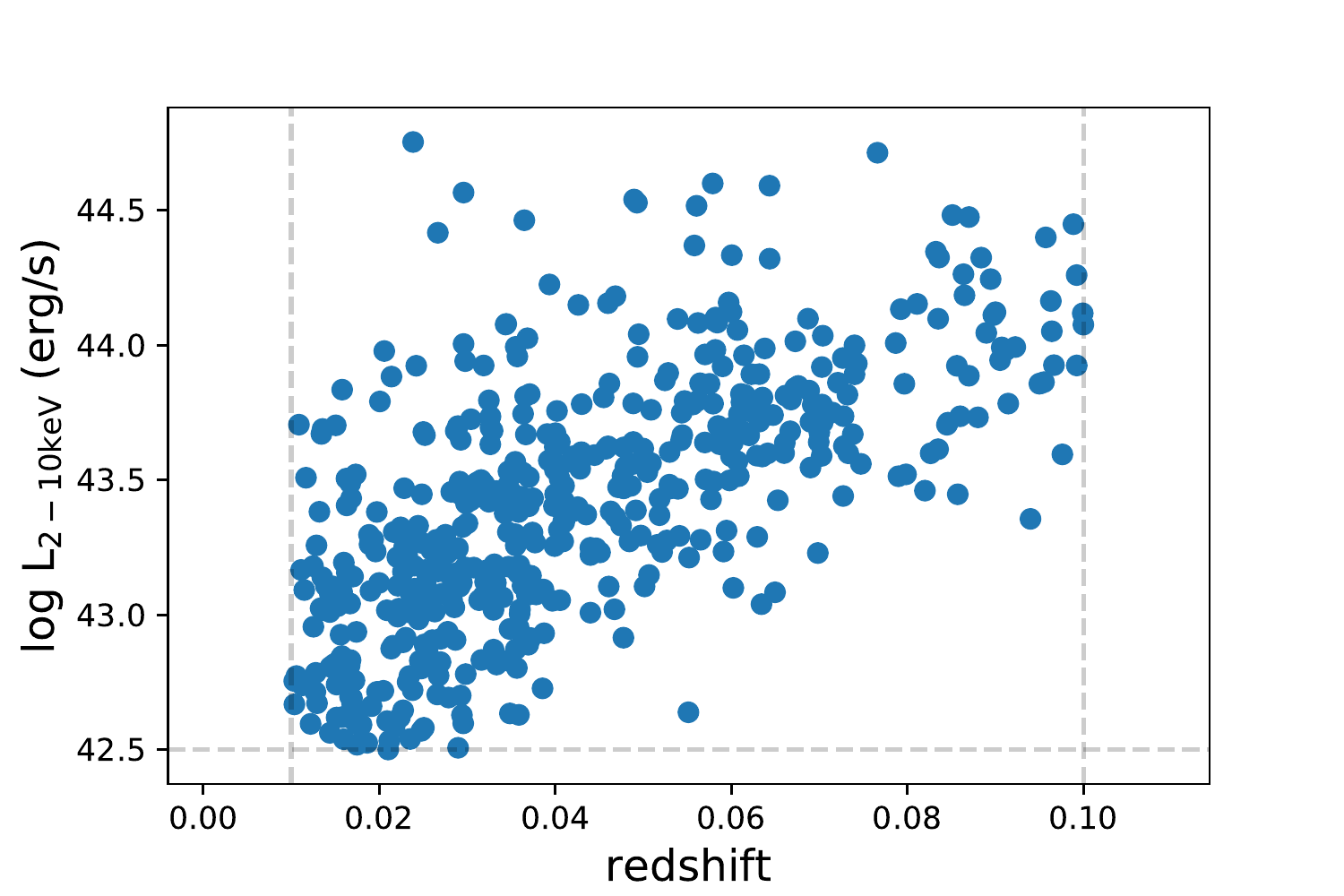}
\caption{Log of the intrinsic $2-10$ keV luminosity versus redshift for the 499 BASS AGN at redshift $0.01<z<0.1$. The sample spans 2 decades in luminosity, but as in all flux-limited samples, there is a strong redshift-luminosity correlation. }
\label{fig:Lz}
\end{figure}

We selected AGN in the redshift range $0.01<z<0.1$ with intrinsic (i.e., absorption-corrected) $L_{2-10~keV}>10^{42.5}$ $\rm erg \,s^{-1}$, to remove any bias in peculiar velocities of low-redshift objects, as well as improve the AGN completeness for the luminosity range. The upper redshift limit was imposed to match the maximum redshift of the galaxy sample that we cross-correlated with the AGN. After these cuts, the final number of AGN in our sample is 499, and their distribution of redshift versus X-ray luminosity is shown in Figure~\ref{fig:Lz}. The luminosities are comparable to those of well-studied, higher redshift AGN from pencil-beam X-ray surveys (e.g., COSMOS; \citealt{Civano:2016,Marchesi:2016}).

In addition to this full sample, we made subsamples in two bins of $N_{\rm{H}}$ (threshold $=10^{22}$ cm$^{-2}$) and two bins of black hole mass (threshold $=10^{8}$ $M_{\odot}$). 
Because the statistics are insufficient to use volume-limited samples, different luminosity subsamples automatically probe different volumes and host galaxy stellar masses. 
We therefore do not examine the clustering dependence on X-ray luminosity. Additionally,
to avoid selection effects between the two bins of $N_{\rm{H}}$ and $M_{\rm{BH}}$, we matched the subsamples in their distributions of redshift and X-ray luminosity. Specifically, we defined five bins of $z$; then for each bin, we randomly selected $N$ AGN of the sample, with the larger number of sources in the bin to match the number of sources ($N$) in the other sample. We then repeated the process for luminosity, with 10 bins of $\log L_{2-10~keV}$. The total numbers were 186 AGN in each bin of $N_{\rm{H}}$ and 102 AGN in each bin of $M_{\rm BH}$. Each random selection provided consistent results, as did using the derived bolometric luminosities rather than $L_X$. The characteristics of these subsamples are summarized in Table \ref{table:dtable}.

 \begin{table}
            
\label{table:dtable}     
\centering                          
\begin{tabular}{c c c c c c}        
\hline   \hline              
 AGN Sample & Threshold & N & $\tilde{M}_{bh}$ & $ \tilde{L}_{2-10~keV}$ & $\langle z\rangle$ \\    
\hline                     
$L$-limited (Full)	&   $L_{X}>10^{42.5}$ erg s$^{-1}$  & 499 & 8.0 & 43.4 & 0.04   \\
$\lambda_{\rm{Edd}}$-limited	 & $ \lambda_{\rm{Edd}}>0.01$ &  245   & 7.9 & 43.5 & 0.04    \\
Obscured 		& $N_{\rm{H}}\geq 10^{22}$ cm$^{-2}$& 186  & 8.2 & 43.4 & 0.04    \\
Unobscured 		& $ N_{\rm{H}}<10^{22}$ cm$^{-2}$& 186 & 7.7 & 43.4 & 0.04     \\
Small $M_{\rm{bh}}$  & $M_{\rm{bh}}\leq 10^{8}~M_{\odot}$ & 102 & $7.6$ & 43.4 &   0.04   \\
Large $M_{\rm{bh}}$	& $M_{\rm{bh}}>10^{8}~M_{\odot}$& 102 & $8.4$& 43.4 &   0.04  \\
      &  &      \\

\hline
\end{tabular}
\caption{AGN subsamples and their characteristics, including the number of AGN, the median black hole mass, the median 2-10 keV luminosity (after correcting for absorption), and the average redshift of each. Black hole mass is in log units of $M_{\odot}$, and luminosity is in log units of erg s$^{-1}$.} 
\end{table}

\subsection{Galaxy Sample}

Using a dense sample of galaxies as tracers of the underlying dark matter distribution greatly boosts AGN clustering statistics \citep[e.g.,][]{Coil:2009}. We therefore cross-correlated our AGN sample with galaxies from the 2MASS Redshift Survey (\citealt{Huchra:2012}), as the redshift range is close to that of the AGN sample ($z_{\rm peak} \sim 0.03$). Selected based on their $K$-band magnitude, $K_{\rm s}\leq 11.75$, the galaxies are spectroscopically complete and cover $91\%$ of the sky (the Galactic plane is excluded; $|b|>8^{\circ}$).

We estimated stellar masses of the 2MASS galaxies by employing a universal mass-to-light ratio ($M/L$) between $K$-band luminosity and stellar mass, as $K$-band $M/L$ does not significantly vary with mass at $z=0$ \citep{Lacey:2008}, nor is it sensitive to dust content. We use an absolute solar $K_{\rm{S}}$ band magnitude of 3.29 \citep{Blanton:2007} to obtain the luminosities and fit our measured autocorrelation function for $M_{*}/L_{K_{S}}$, as described in section 3.3. The random error associated with using a single $M/L$ ratio is about 0.3 dex \citep{BelldeJong:2001}. However, we only use the resulting distribution of stellar mass in our model, and we verified that convolving the distribution with this error does not change our results.

We used the full flux-limited sample for maximal statistics, and corrected for incompleteness as a function of stellar mass when modeling the galaxies via the process described in Section 3. We excluded 2MASS galaxies that are also in the BASS AGN catalog (to within $3^{\prime\prime}$; 361 sources) so that the cross-correlation measurement was between two independent catalogs. In total, we used 38,567 galaxies in the redshift range $0.01<z<0.1$.

\section{Method}
\subsection{Correlation Function Measurement}

The quantitative measure of clustering is the two-point correlation function, which quantifies the excess probability over a random distribution that a pair of objects are separated by a given distance ($\vec{r}$).
We used the Landy$-$Szalay estimator \citep{Landy:1993}:

\begin{equation}
\xi(\vec{r}) = \frac{D_{1}D_{2}(\vec{r})-D_{1}R_{2}(\vec{r}) - R_{1}D_{2}(\vec{r}) + R_{1}R_{2}(\vec{r})}{R_{1}R_{2}(\vec{r})}  ,
\end{equation}
where DD, DR, and RR correspond to the data$-$data, data$-$random and random$-$random pairs, respectively. For an autocorrelation (ACF) measurement, the subscripts correspond to the same dataset, while they represent two different datasets for a cross-correlation. The random catalogs for each dataset have the same selection function as the data survey. Rather than using the AGN ACF, which has large uncertainties because of the rather small AGN sample, we cross-correlated the AGN with the larger galaxy sample to improve statistics.

We created a random AGN sample with the same selection as the BASS survey by using the \emph{Swift}/BAT sensitivity map. We first randomized the the position of each random AGN on the sky, and then assigned it a flux drawn from the flux distribution of the data. If the flux was greater than the sensitivity at that position, we kept that specific randomly generated AGN; otherwise we omitted it. We then assigned it a redshift drawn from the redshift distribution of the data, smoothed with a Gaussian kernel with $\sigma_{z}=0.2$. We repeated this process for each AGN subsample (e.g., each bin in black hole mass or in absorbing column density). Due to the low number density of the data, we made each random AGN sample $\sim 100$ times larger than the corresponding BASS sample. 

\begin{figure}
\centering
\includegraphics[width=.5\textwidth]{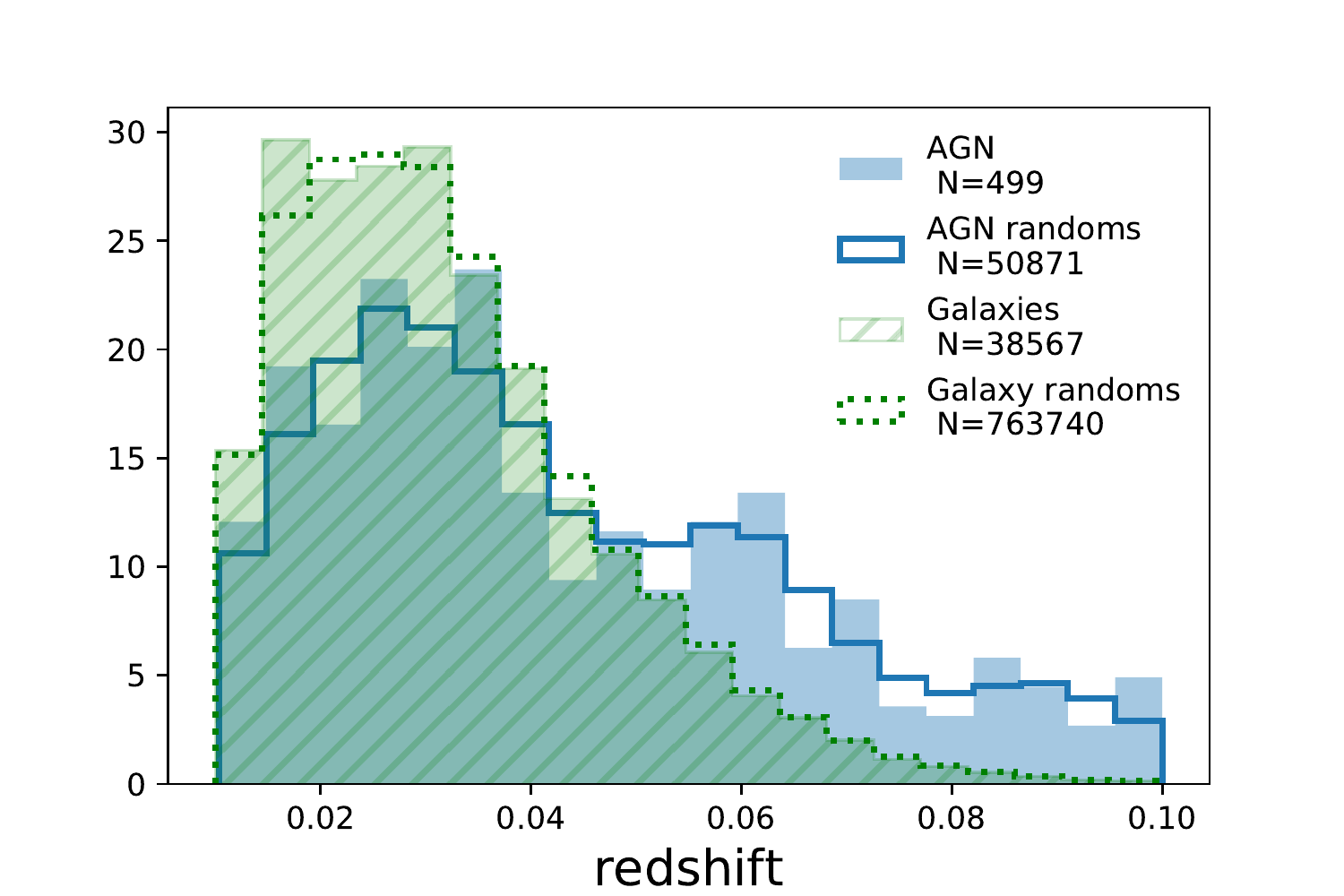}
\caption{Normalized redshift distributions of the AGN and galaxy samples, along with their respective random catalogs. The redshift distribution of each population is well matched by its randomly positioned counterparts.}
\label{fig:zdist}
\end{figure}

For the galaxy random catalog, we assumed that the sensitivity is uniform across the sky and randomized the angular positions, excluding the galactic plane ($|b| < 8^{\circ}$). We assigned each random galaxy a redshift drawn from the distribution in the real data, also smoothed with a Gaussian kernel with $\sigma_{z}=0.2$. The redshift distributions of the galaxies, AGN, and random samples are shown in Figure \ref{fig:zdist}. The number of random galaxies is 20 times more than the number of 2MASS Redshift Survey galaxies.

We measured $\xi$ in bins of $r_{\rm{p}}$ and $\pi$ (distances perpendicular and parallel to the line of sight, respectively) using the pair counter from the publicly available software \texttt{CorrFunc} \citep{Sinha:2017}, which counts the number of pairs of galaxies in a catalog separated by $r_{\rm{p}}$ and $\pi$. We then projected through redshift space to eliminate any redshift-space distortions, to get the projected correlation function:

\begin{equation}
w_{p} = 2 \int^{\pi_{\rm{max}}}_{0} \xi(r_{p},\pi) d\pi   .
\end{equation}
The value of $\pi_{\rm{max}}$ was chosen such that the amplitude of the projected correlation function converges and gets noisier for any higher values. We found this to be $60$ Mpc $h^{-1}$ for our sample, which is a commonly used value for $\pi_{\rm{max}}$.

We calculated the covariance matrix via the jackknife resampling method:

\begin{equation}
\begin{split}
C_{i,j} = \frac{M}{M-1} \sum_{k}^{M} \Big[w_{k}(r_{p,i}) - \langle w(r_{p,i})\rangle\Big]\\
\times\Big[w_{k}(r_{p,j}) - \langle w(r_{p,j})\rangle \Big] ~~,
\end{split}
\end{equation}

\noindent
where we split the sample into $M=25$ sections of the sky, and computed the cross-correlation function when excluding each section ($w_{k}$). We chose $M=25$ so that the patches were large enough to probe the largest $r_{\rm{p}}$ scale at our minimum redshift, yet numerous enough to create a normal distribution. We quote the errors on our measurement as the square root of the diagonals: $\sigma_{i} = \sqrt{C_{i,i}}$.

\subsection{Model Formulation}

In the hierarchical model of structure formation, galaxies reside in dark matter halos, which have gravitationally collapsed at the peaks of the underlying dark matter distribution. In this context, clustering statistics of galaxies depend only on the cosmology (how dark matter halos cluster; the two-halo term, dominant on scales $\gtrsim$1 Mpc h$^{-1}$) and how the galaxies occupy their dark matter halos (one-halo term; $\lesssim$1 Mpc h$^{-1}$), which depends on their formation and evolution. We consider two kinds of models to describe the latter: a HOD model and a subhalo model, described in the following sections. 

In both cases, we used \texttt{Halotools} \citep{Hearin:2017} to compute the model cross-correlation functions. This software populates dark matter halos from an $N$-body simulation with a model and computes the two-point statistics for the resulting galaxy mock catalog. Because we cross-correlated AGN with galaxies, we first created a mock sample with the same clustering statistics as the 2MASS galaxies (described in Section 3.3). We then used this simulated galaxy sample to cross-correlate with the AGN mock derived from the model. The average and median halo masses of the AGN sample were calculated empirically from the AGN mock.

We did the analysis with two redshift $z=0$ halo catalogs (based on the ROCKSTAR halo-finder; \citealt{Behroozi:2013}) from different simulations, both included in \texttt{Halotools}: first, the Bolshoi--Planck simulation \citep{Riebe:2011}, which has a resolution of $1.35\times 10^8 M_{\odot}h^{-1}$ and a box size $L=$250~Mpc h$^{-1}$ using Planck 2013 cosmological parameters \citep{Planck:2015}; second, the Consuelo simulation, which has a larger volume ($L=420$~Mpc h$^{-1}$) but poorer resolution ($2\times 10^9 M_{\odot}h^{-1}$), with WMAP5 cosmology \citep{wmap5:2009}. The results are consistent with each other; however, since the Bolshoi--Planck simulation is complete down to haloes of mass $M_{\rm{vir}}\sim 10^{11} M_{\odot}$, we quote results from that analysis, which is better able to constrain minimum halo mass.

\subsubsection{HOD Model}
The HOD formalism \citep[e.g.,][]{Cooray:2002} describes the probability that $N$ galaxies (or AGN) reside in a host halo of mass $M_{h}$. To first order, this can be described as the average number of galaxies per host halo as a function of halo mass, $\langle N\rangle(M_{h})$. 

The HOD can be disaggregated into centrals and satellite galaxies, where the total HOD is sum of the two components $\langle N\rangle = \langle N_{c}\rangle + \langle N_{s}\rangle$. We used a simple parametrization for the the AGN HOD, derived from the \cite{Zheng:2007} model:
\begin{equation}
\langle N_{c}\rangle(M_{\rm{h}}) \propto \Theta( M_{h} - M_{\rm{min}})~,
\end{equation}

\begin{equation}
\langle N_{s}\rangle(M_{\rm{h}}) \propto \Big(\frac{M_{\rm{h}}-M_{\rm{min}}}{M_{1}}\Big)^{\alpha} ,
\end{equation}

\noindent

where $\Theta$ is the Heaviside step function, $M_{\rm{min}}$ is the minimum halo mass to host a central AGN, $M_{1}$ is typical halo mass that starts hosting satellites, and $\alpha$ is the power-law slope of the satellites.
We assumed $~\log(M_{1}/M_{\rm{min}})=1.2$, which is the case for galaxies with $M_r<-20$~mag \citep{Zehavi:2011}, and we left $M_{\rm{min}}$ and $\alpha$ as two free model parameters. The normalization of the HOD is  not constrained by the correlation function. 
We searched for the best-fit model by stepping through $\log M_{\rm{min}}-\alpha$ parameter space in 0.1 unit increments ($11.2<\log M_{\rm{min}}<12.8$; $-0.5<\alpha<1.5$), where at each step we averaged five model realizations, and found where the correlated $\chi^{2}$ was minimum:

\begin{equation}
\begin{split}
\chi^{2} = \sum_{i,j} \Big[ w_{\rm{obs}}(r_{\rm{p},i}) - w_{\rm{mod}}(r_{\rm{p},i})\Big]
\times~C^{-1}_{\rm{eff},i,j}\\
\times\Big[w_{\rm{obs}}(r_{\rm{p},j}) - w_{\rm{mod}}(r_{\rm{p},j})\Big]  ~,
\end{split}
\end{equation}

\noindent
where $w_{\rm{obs}}$ and $w_{\rm{mod}}$ correspond to the correlation function of the real data and mock data. 
Because the model has sample variance uncertainty from the finite simulation volume, $C_{\rm eff}$ is the sum of the covariance matrices from the data and simulation \citep{Zheng:2016}. 
The simulation covariance matrix was also estimated via Jackknife resampling, by splitting the simulation box into 125 cubes.
We report the best-fit parameters in Section 4. 

For each realization of the HOD model, \texttt{Halotools} populates the the host halos with the mock central galaxies and adds satellites according to an NFW profile \citep{NFW:1996}. This is done only for the largest virialized halos in the catalog (i.e., ignoring the subhalo information).

\subsubsection{Subhalo Model}
The second type of model assumes a one-to-one relation between the galaxies and all halos and subhalos.
We used the \cite{Behroozi:2010} model based on abundance matching, which assumes that stellar mass predominantly determines the clustering of the sample via the stellar-to-(sub)halo mass relation. This model has been calibrated and tested with galaxy observations, and so it provides an additional check to see if AGN occupy halos in the same way as inactive galaxies,  i.e., based primarily on their stellar mass.

For this model, the \texttt{Halotools} software populates a mock galaxy at the center of each halo {\it and} subhalo, and assigns it a stellar mass based on the peak mass of that (sub)halo. The mock galaxies in the  center of each host halo correspond to the centrals galaxies, while the mocks in the subhalos correspond to the satellites.

This method allows us to correct for the incompleteness of the AGN catalog in the following way: we populated the halos from our halo catalog with galaxies according to the Behroozi model, and then divided the stellar mass distribution of the resulting galaxy mock sample with the stellar mass distribution of the BASS AGN. We normalized it to obtain the incompleteness fraction as a function of stellar mass. We then assigned random values between 0 and 1 to the mock galaxies, and masked out any mock whose value fell below the incompleteness fraction for its assigned stellar mass. Consequently, the resulting mock sample had the same stellar mass distribution as the data.

This subhalo-based model is approximately equivalent to the HOD model that assumes $\alpha=1$, as the number of subhalos (and hence satellite galaxies above a threshold luminosity) scales with halo mass. However, it is not biased by incompleteness in flux-limited catalogs. For this model there are no parameters to fit; rather, we simply assess how well the model agrees with the data.

\subsection{Galaxy Mock Creation}

\begin{figure}
\centering
\includegraphics[width=.5\textwidth]{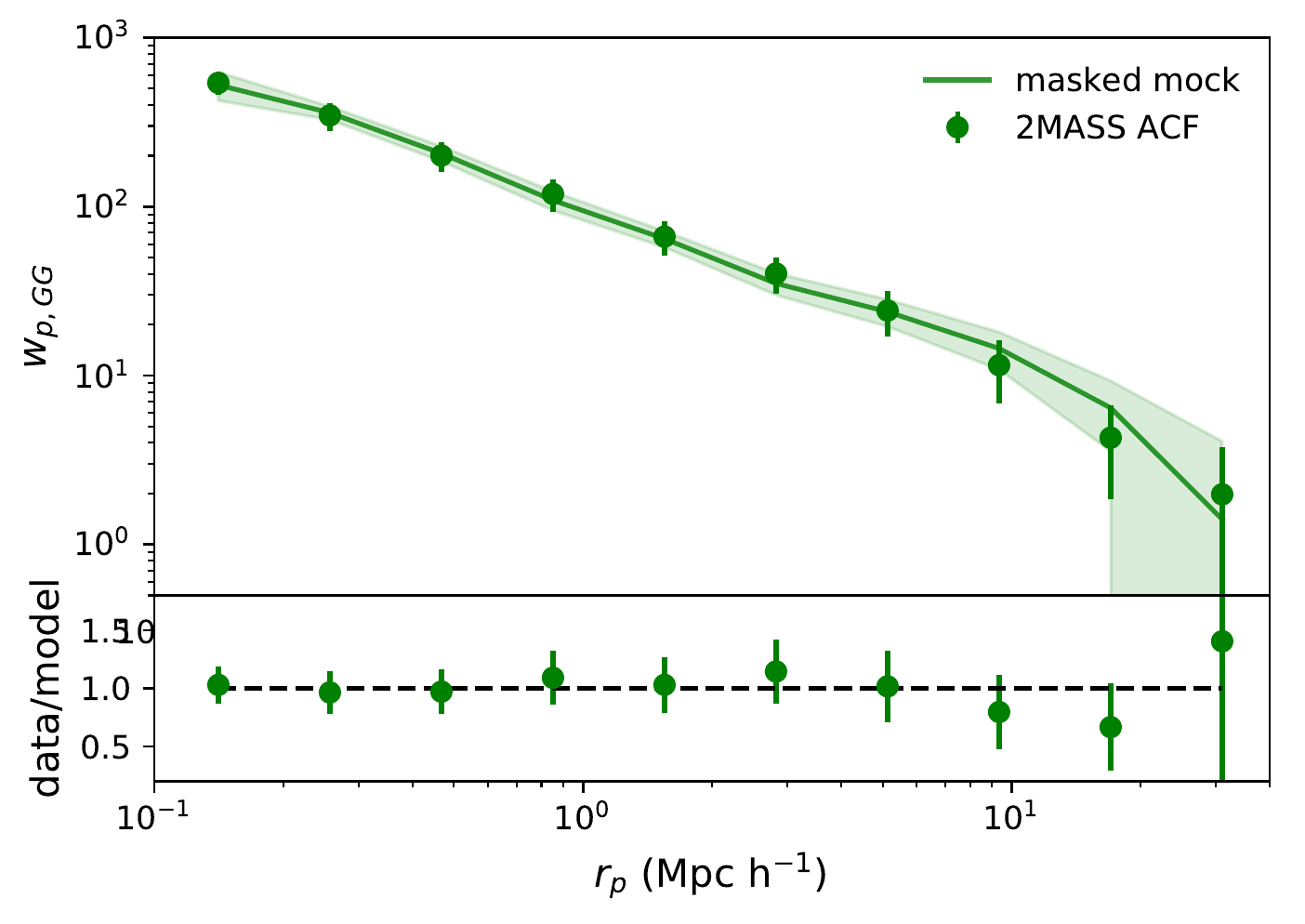}
\caption{Projected autocorrelation function of 2MASS galaxies compared with the mock sample created using the \cite{Behroozi:2010} model and 2MASS selection function. 
}
\label{fig:wp_gg}
\end{figure}

We used the subhalo model to create a mock galaxy sample with the same stellar mass distribution as the full flux-limited 2MASS galaxy catalog. We fit for the $K_{s}$-band $M/L$, by comparing the resulting mock autocorrelation function with the data. The best-fit value was found to be 0.6 in solar units ($\chi^{2}_{\nu}=0.75$).
The masked autocorrelation function using the Bolshoi--Planck halo catalog with the best-fit $M/L$ ratio is shown in Figure \ref{fig:wp_gg}, along with the autocorrelation function of the 2MASS galaxies. We found that using $M/L$ ratios of upper and lower bounds of the 99\% confidence limits of the fit does not significantly change our results. Both simulations produced consistent results.

\section{Results}

The results for both models are summarized in Table 2 for each AGN sample.

\subsection{Full AGN Sample}

\begin{figure*}
\centering
\includegraphics[width=.48\textwidth]{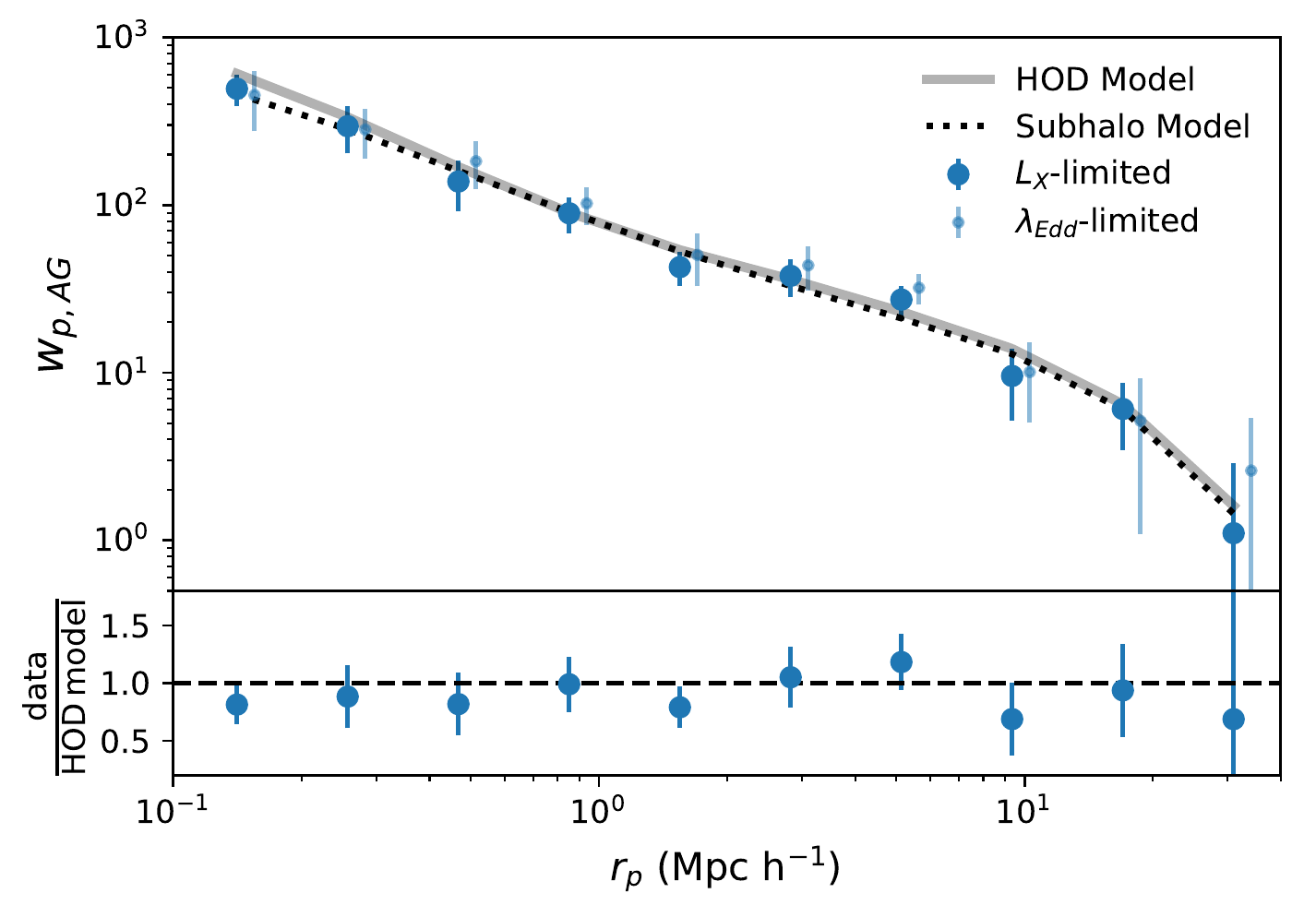}
\includegraphics[width=.48\textwidth]{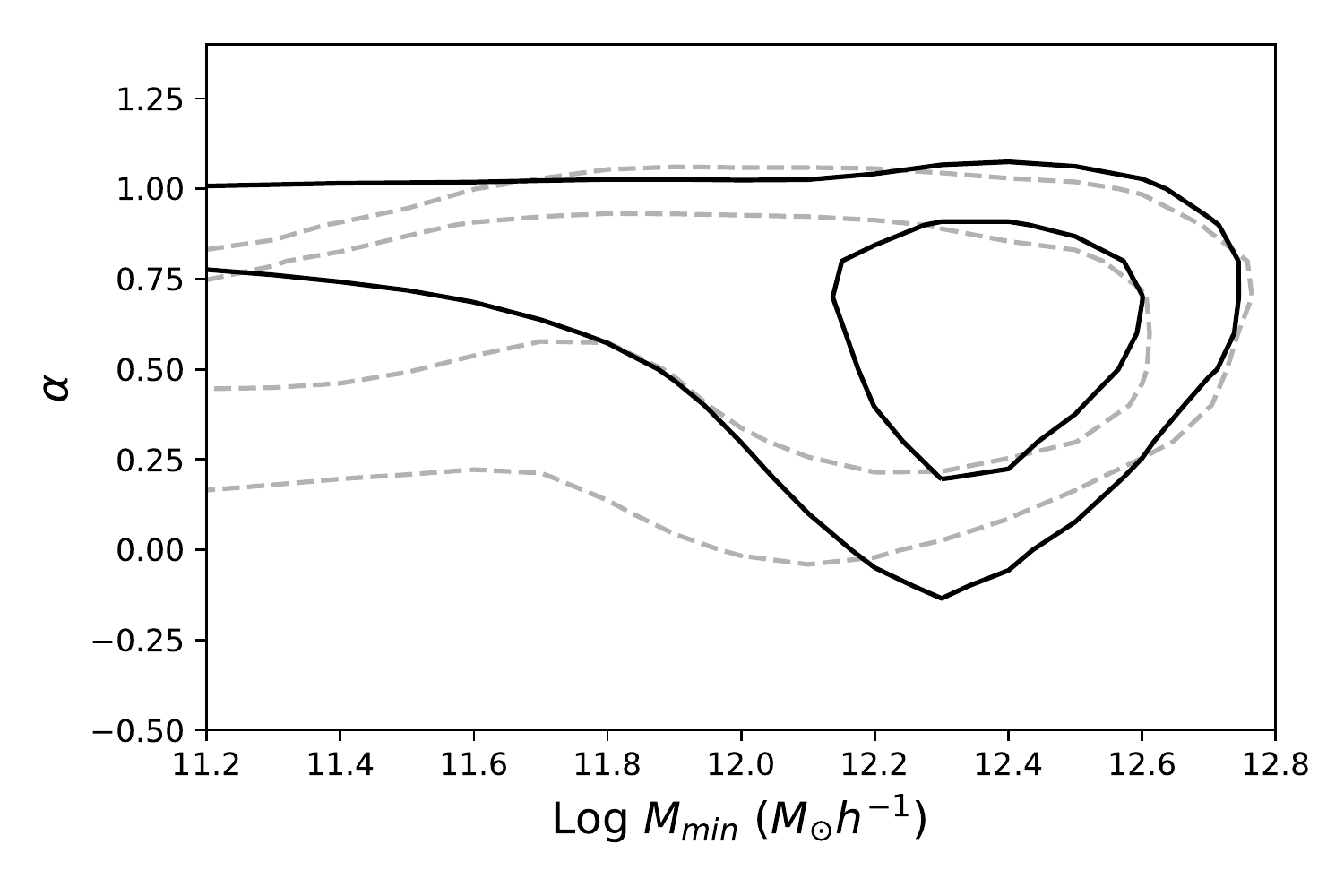}
\caption{Left: projected cross-correlation function of 2MASS galaxies and BASS AGN (blue points), with the best-fit HOD model (gray solid line) and subhalo model (black dotted line) for the $2-10$ keV luminosity-limited sample. The lower panel shows the data divided by the HOD model. An AGN sample limited by Eddington ratio ($L/L_{\rm{Edd}} > 0.01$; light blue points) is consistent with the same models. Right: contour map of the HOD fit, showing the $\Delta \chi^{2}=2.3$ and 6.2 levels, for the Bolshoi--Planck catalog (solid lines) and the Consuelo halo catalog (dotted lines).}
\label{fig:wp_all}
\end{figure*}

The left panel of Figure \ref{fig:wp_all} shows the projected cross-correlation function for the full AGN sample and the corresponding HOD model fit: $\log M_{\rm{min}}/M_{\odot}h^{-1} = 12.4^{+0.2}_{-0.3}$, $\alpha_{AGN}=0.8^{+0.2}_{-0.5}$. We find that the average dark matter halo mass in which AGN reside is $\log M_{h}/M_{\odot}h^{-1}=13.4\pm0.2$, and the median mass is $\log M_{h}/M_{\odot}h^{-1} = 12.8\pm0.2$, from empirical measurement of the mocks from the $1\sigma$ best-fit HOD region. This is consistent with the measurement in \cite{Cappelluti:2010}, as well as in \cite{Krumpe:2017}. The smoothed contour map of the fit to the two-parameter HOD is shown in the right panel. While the associated significances of the contour levels should be taken with caution, we verified that the projected probability distributions for each parameter are nearly Gaussian, which we use to quote the errors on the best-fit parameters. 
The satellite power-law slope is consistent with that of the local inactive galaxy population ($\alpha \sim 1$), but favors $\alpha<1$. We stress that this HOD fit is using the full flux-limited sample, which is incomplete for all AGN luminosities. Thus the derived HOD pertains to AGN with median bolometric luminosity of $10^{44.7}$~erg~s$^{-1}$ at $z\sim 0.04$. However, we are able to compare our full cross-correlation measurement with an Eddington ratio-limited sample ($\lambda_{\rm{Edd}}>0.01$), which has been suggested to have a more unbiased HOD \citep{Jones:2017} than a luminosity-limited sample. Although the statistics are poorer due to only a fraction of the sources having black hole mass estimates, we find it also agrees with our best-fit HOD model (Figure \ref{fig:wp_all}).

Figure \ref{fig:wp_all} also shows the results from our subhalo model analysis, which agrees well  with the data ($\chi^{2}_{\nu}=1.6$), despite there being no free parameters. The advantage of the subhalo model is that it takes into account the incompleteness of the catalog. The median halo mass, $\log~M_{h}/M_{\odot}h^{-1}\sim 12.3$, is lower than was found with the HOD model (12.8) because of the proper treatment of incompleteness (i.e., taking into account the smaller mass galaxies and halos that were missed).
We therefore conclude that AGN, on average, do not live in special environments compared with the overall galaxy population, as our only assumption was that the host galaxy stellar mass distribution of the AGN sample drives its clustering via the stellar-to-subhalo mass relation.

\begin{table*}
         
\label{table:t2}      
\centering     

\begin{tabular}{c | c c c c c | c c c}        
 & &  & HOD Model & & & & Subhalo Model & \\
\hline   
 AGN Sample & $\tilde{M_{h}}$ & $\langle M_{h}\rangle$ & $M_{min}$ & $\alpha$ &  $\chi^{2}_{\nu}$ &  $\tilde{M_{h}}$& $\langle M_{h}\rangle$ & $\chi^{2}_{\nu}$ \\    
\hline     \hline   
Full &  $12.8^{+0.2}_{-0.1}$ & $13.4^{+0.1}_{-0.3}$ & $12.4^{+0.2}_{-0.3}$ & $0.8^{+0.2}_{-0.5}$ & 1.5  & 12.3 & 13.3 & 1.6\\

Obscured & $12.9^{+0.3}_{-0.7}$ & $13.5^{+0.2}_{-0.2}$ & $12.5^{+0.2}_{-0.8}$&$1.1^{+0.4}_{-0.2}$ & 1.9 & 12.3&13.3 & 2.1\\

Unobscured  & $12.0^{+0.2}_{-0.3}$  & $12.6^{+0.2}_{-0.3}$  & $11.4\pm0.2$  & $0.4^{+0.2}_{-0.4}$&1.0  &12.3 &13.3& 4.5\\

Small $M_{\rm{bh}}$  & $12.6^{+0.2}_{-0.9}$ & $13.4^{+0.2}_{-0.9}$ & $12.1^{+0.4}_{-1.0}$ & $0.9^{+0.2}_{-0.4}$ & 2.6 & 12.2&13.3 &2.1\\

Large $M_{\rm{bh}}$  & $12.8^{+0.2}_{-0.4}$ & $13.2^{+0.2}_{-0.3}$ & $12.4^{+0.2}_{-0.4}$ & $0.2^{+0.5}_{-0.4}$ & 0.6 & 12.4&13.3 &1.6\\

\hline

\end{tabular}
\caption{Halo model parameters for each AGN subsample, for both the HOD and subhalo models. All masses are in log units of $M_{\odot} h^{-1}$.}  
\end{table*}

\subsection{Clustering versus Obscuration}

\begin{figure*}
\centering
\includegraphics[width=.49\textwidth]{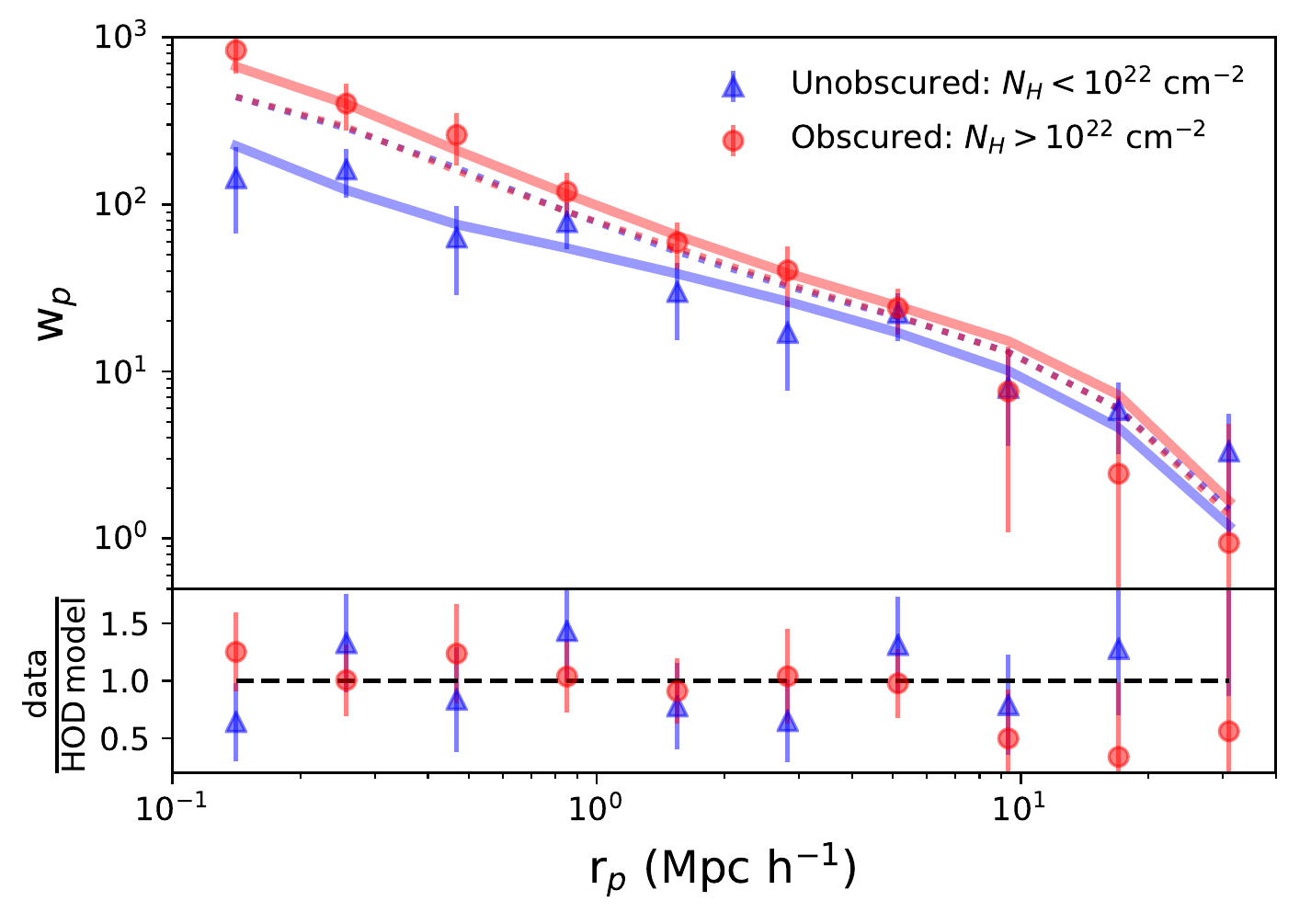}
\includegraphics[width=.49\textwidth]{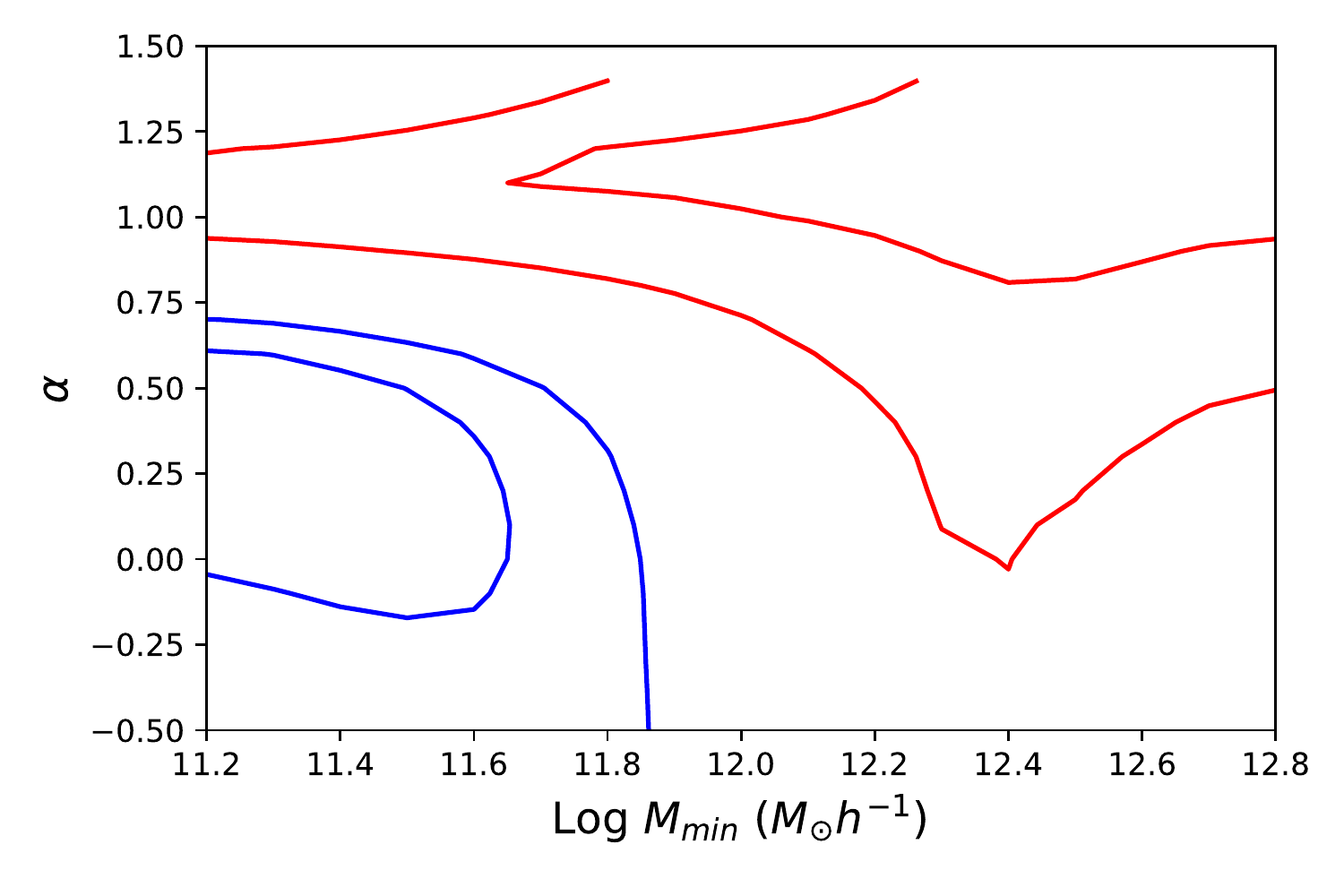}
\includegraphics[width=.75\textwidth]{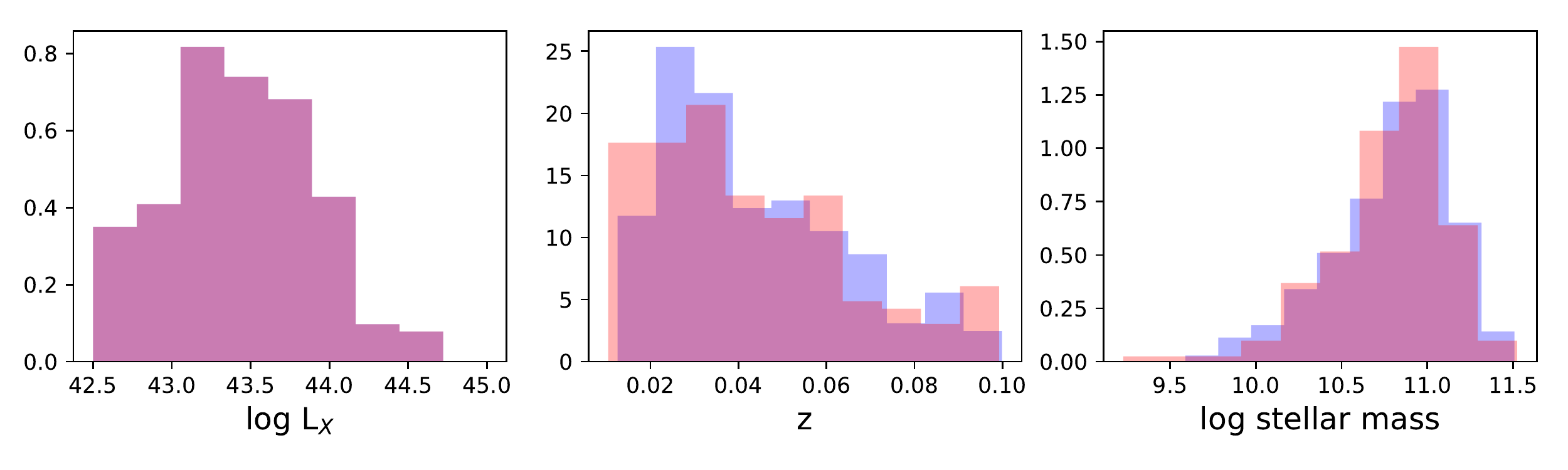}
\caption{Upper panels: projected cross-correlation function of obscured (red) versus unobscured (blue) AGN and corresponding HOD model fits (solid lines) and subhalo models (dotted points). While their two-halo terms are consistent with each other, obscured AGN appear more clustered on scales of the one-halo term. Upper right: $\Delta \chi^{2}$ contour maps of the HOD fit for unobscured (blue) and obscured (red) AGN are completely distinct, suggesting some difference between the two populations. Lower panels: matched subsamples have similar distributions of (left) log of the $L_{2-10~keV}$ luminosity, (middle) redshift, and  (right) log of the host galaxy stellar mass.}
\label{fig:nH}
\end{figure*}

Figure \ref{fig:nH} shows the cross-correlation function of unabsorbed ($N_{\rm{H}}<10^{22}$ cm$^{-2}$) versus absorbed ($N_{\rm{H}}\geq 10^{22}$ cm$^{-2}$) AGN with their corresponding HOD fits. The luminosity and redshift distributions of the two bins are shown. 

While the two-halo terms of the data seem consistent with each other, the obscured AGN appear more clustered on small scales (by $\sim 3\sigma$), consistent with recent studies of narrow- versus broad-line AGN in the Sloan Digital Sky Survey (SDSS; \citealt{Jiang:2016}) and in {\it Swift}/BAT AGN \citep{Krumpe:2017}. This was also seen using the full sample (without matching Luminosity distributions) and using different $N_{\rm{H}}$ thresholds up to (but not including) $10^{23}$ cm$^{-2}$.

The stellar mass distributions shown in Figure \ref{fig:nH} are very similar, so this cannot cause the difference in clustering. The subhalo model for unobscured AGN (dotted blue line) is inconsistent with the data (blue dots)
($\chi^{2}_{\nu} > 4$), another indication that factors beyond host galaxy stellar mass are determining the clustering signal. 

The $\Delta \chi^{2}$ contour plots of the separate HOD fits for obscured and unobscured AGN are also shown in Figure \ref{fig:nH}; the shapes of the HODs differ by more than $4\sigma$. $M_{min}$ and $\alpha$ are different for each: unobscured AGN have smaller minimum halo mass and a shallower satellite-term slope. This would suggest that unobscured AGN tend to be in central galaxies while obscured AGN are more likely to be in satellites. The corresponding average dark matter halo masses are $\log M_{h}/M_{\odot}h^{-1}=13.5\pm 0.2$ for obscured AGN and $\log M_{h}/M_{\odot}h^{-1}=12.6\pm 0.3$ for unobscured AGN. The finding that obscured AGN live in larger mass halos than their unobscured counterparts agrees with recent results of angular clustering studies of infrared-selected \emph{WISE} AGN \citep{Hickox:2009,DiPompeo:2014,DiPompeo:2017}. It is inconsistent, however, with clustering studies of Type 1 vs. Type 2 X-ray-selected AGN in COSMOS \citep{Allevato:2014}, although these studies probed AGN at higher redshift ($z\sim 1$) and different luminosity ranges.
This inconsistency could also be due to the host galaxies; the Type 1 sample had systematically higher luminosities, indicating they most likely had larger host galaxy stellar masses, which may explain why a larger bias for Type 1 AGN was found.
The clustering properties of unobscured AGN are also consistent with the halo masses found for optical quasar samples across a wide range of redshift \cite[e.g.,][]{Croom:2005,Ross:2009,Shen:2009}.

\begin{figure}
\centering
\includegraphics[width=.5\textwidth]{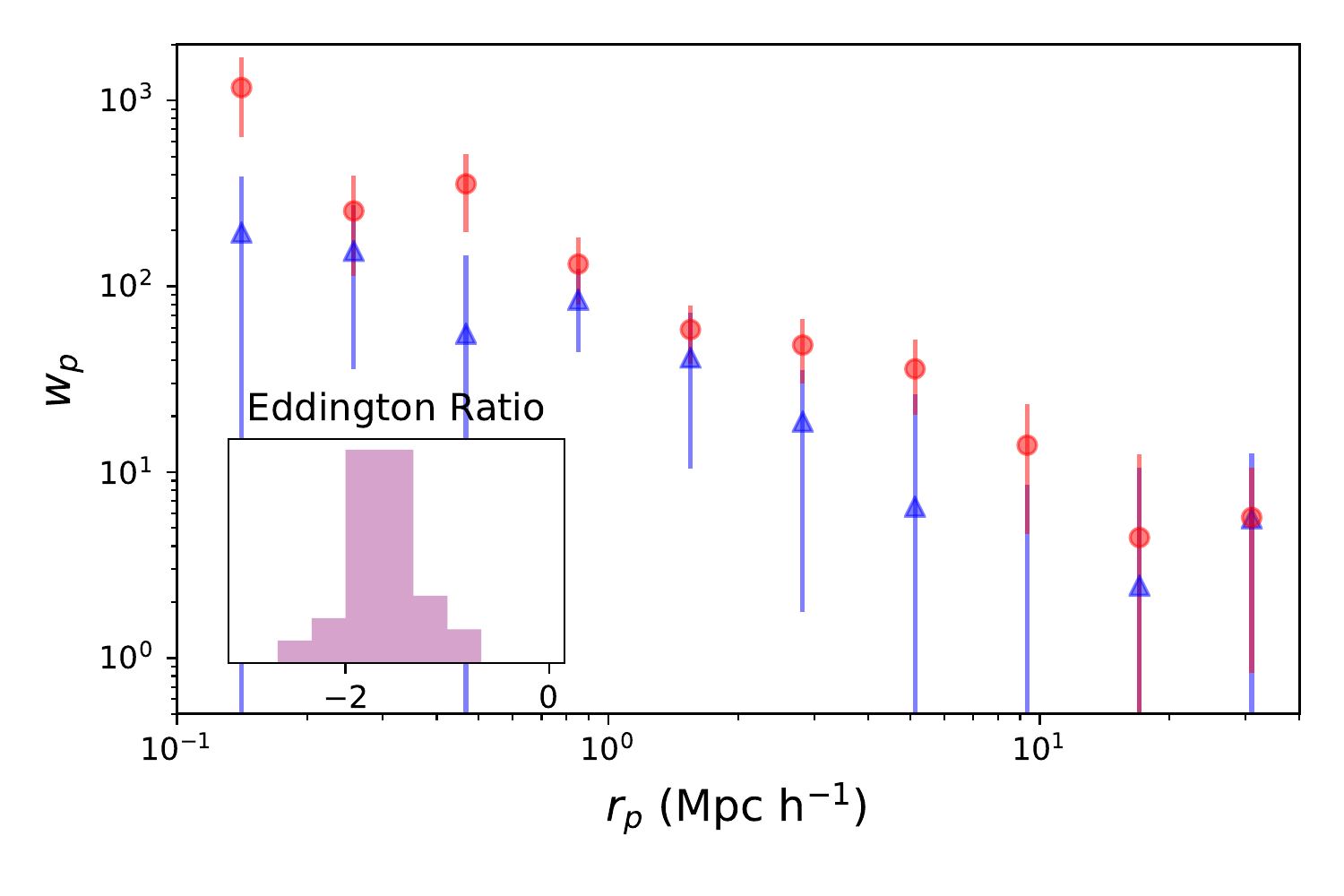}
\caption{Projected cross-correlation function for Compton-thin obscured (red) and unobscured (blue) AGN with matched distributions of Eddington ratio (inset). Although this analysis involved only half as many AGN as in Figure 4, the difference is similar, indicating that black hole mass (and its possible relation to halo mass) is not causing the difference in clustering.}
\label{fig:t1t2_edd}
\end{figure}

The distinctly different halo masses of obscured and unobscured AGN could be due to intrinsic differences between the two types. It has been suggested that (Compton-thin) obscured AGN tend to have lower Eddington ratios than unobscured AGN, since the covering factor depends on mass-normalized accretion rate \citep[e.g.,][]{Ricci:2017}. This would cause our sample of obscured AGN to have systematically larger black holes than the unobscured AGN since we matched their luminosities, which we verified with their $M_{\rm BH}$ distributions. To test if this is biasing the result, we considered the objects that have black hole mass and accretion rate estimates ($\sim 75\%$ of the AGN sample analyzed), and measured the clustering of Compton-thin ($N_{\rm H}<10^{23.5}$ cm$^{-2}$) obscured and unobscured AGN after matching distributions of Eddington ratio rather than luminosity (Figure \ref{fig:t1t2_edd}). The differences between the two types are still present, suggesting something else determines the environmental differences. However, it should be noted that the black hole mass determination for obscured AGN is less precise than for unobscured AGN.

\begin{figure}
\centering
\includegraphics[width=.5\textwidth]{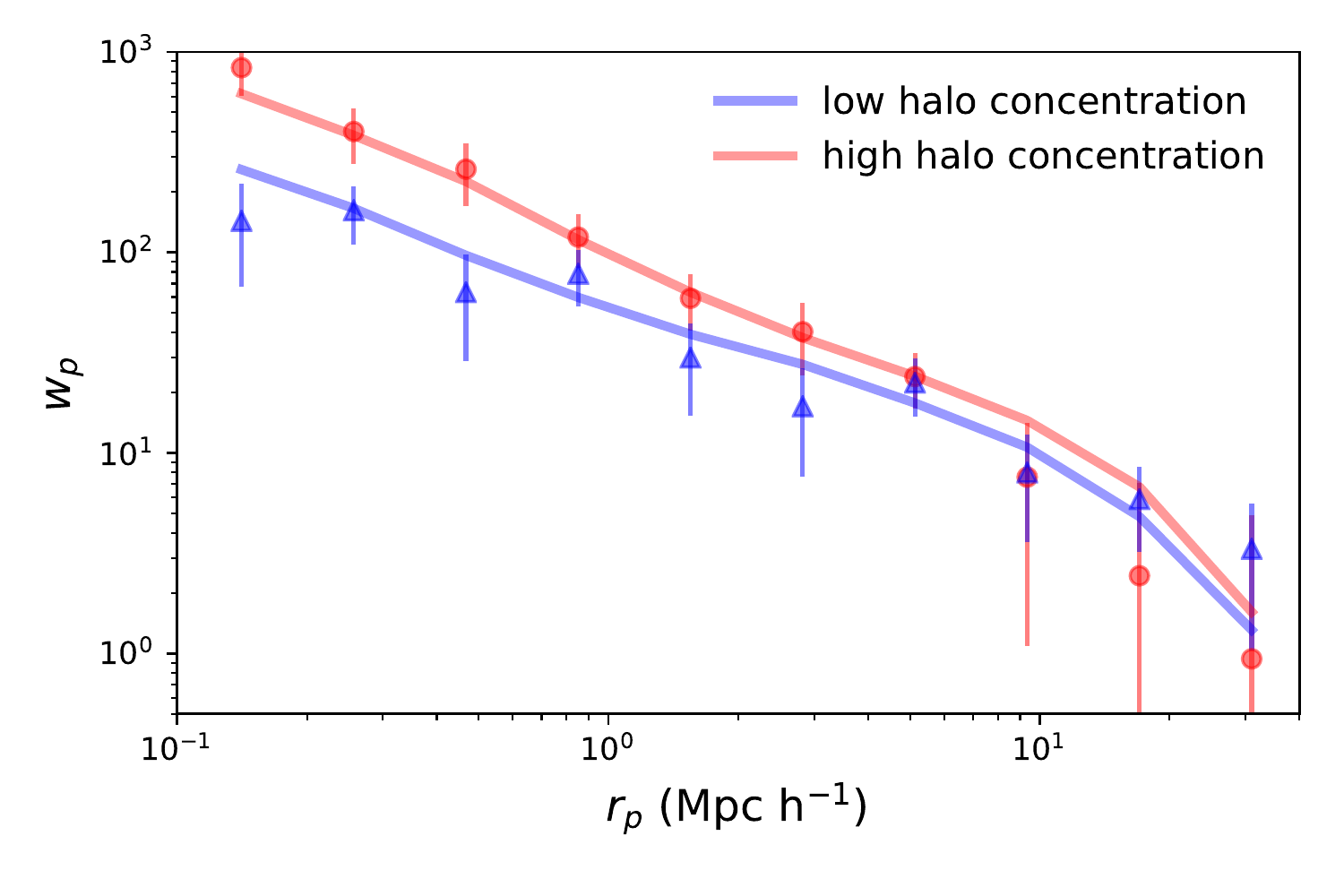}
\caption{The clustering of obscured (red) and unobscured (blue) AGN subsamples is well reproduced by a toy subhalo model split by host halo concentration ($c<13.5$, blue; $c>10.0$, red). This differs from the HOD interpretation that each population has distinct occupation statistics; rather, each population could reside in halos of statistically different concentrations (and hence different assembly histories).}
\label{fig:conc}
\end{figure}

\subsubsection{Role of Assembly Bias}

Another possible difference of clustering between obscured and unobscured AGN could be related to the host halo assembly history rather than the total halo mass, an affect known as assembly bias \citep[e.g.,][]{Gao:2005,Dalal:2008}. In general, there could be a connection between the mass assembly onto the host halo and the mass assembly onto the central black hole, and additionally, a link between obscuration and halo age can constrain whether obscuration is an evolutionary AGN phase. Also, if mergers are significant for AGN obscuration, it is possible that the merging of the subhalos, relating to the assembly of the host halo, would leave an imprint on the clustering signal.
It has been shown that the amount of substructure in halos of a given mass depends on formation epoch \citep[e.g.,][]{Gao:2005,Dalal:2008}, as subhalos in early-forming hosts have more time to fall toward the center and thus are more concentrated. 
Therefore, if unobscured AGN reside preferentially in halos formed late, then both the one-halo and two-halo terms of their correlation function would be reduced. 
This is quite a different explanation than suggested by the HOD analysis (that they are preferentially central galaxies). 

We investigated this scenario with a simple toy model: we populated the halo catalog with our subhalo model and then split the sample by the NFW concentration of their host halos ($c\equiv r_{\rm vir}/r_{s}$, where $r_{\rm vir}$ is  the virial radius and $r_{s}$ is the NFW scale radius), which correlates with halo formation time \citep[e.g.,][]{Wechsler:2002}. We assumed there is a maximum threshold concentration to host an unobscured AGN, and a minimum threshold concentration to host an obscured AGN. While reality is likely to be more complicated, this simple model can explain the overall trend. We found that the obscured sample is best fit by $c>10\pm 2$ ($\chi^2_{\nu}=1.8$), and the unobscured sample is best fit by $c<13\pm 2$ ($\chi^2_{\nu}=0.9$; the median concentration of the mock sample is $c\sim 10$). Figure \ref{fig:conc} shows the projected cross-correlation function of both models, compared to the obscured and unobscured AGN samples, and Table \ref{table:ctable} summarizes the best-fit parameters. There is good agreement with the data, even with such a simple model. The average halo concentrations of each mock sample are $c\sim 8.5$ (unobscured) and $c\sim 27$ (obscured), which correspond to the formation epochs of $z\sim 1$ and $z\sim 5.5$, respectively (see \citealt{Wechsler:2002}, who define the formation epoch as the time when the halo mass accretion rate, $d\log M_{h}/d\log a$, falls below 2).

From this exercise, we see that obscured AGN do not necessarily reside in more massive halos than unobscured AGN (concentration is inversely proportional to mass); rather, it is possible that unobscured AGN instead prefer halos with low concentration and/or later formation epochs. Evidence of this was seen in a comparison between narrow- and broad-line AGN in SDSS; Type 2 AGN seem to reside in groups that are more centrally concentrated \citep{Jiang:2016}. It remains unclear whether the distribution of satellites or the halo formation epoch would be driving this preference. Note that these results are the opposite of what we would expect for the evolutionary picture in which a merger-triggered AGN is first obscured and then evolves into an unobscured phase; in that case, the obscured AGN would reside in the most recently formed halos, with much smaller difference in average host halo formation epoch between each sample.

 \begin{table}     
\centering                          
\begin{tabular}{c c c c c c c c}        
\hline   \hline              
 AGN Sample & $c_{min}$& $c_{max}$& $\langle c\rangle$   & $\tilde{M_{h}}$ & $ \langle M_{h} \rangle$ & $\chi^{2}_{\nu}$ \\    
\hline                     
Obscured 	& $10.0^{+1.5}_{-2.0}$ &	- &8.5&  $12.3$ & $13.4$ & 1.8     \\
Unobscured 	& - & $13.5^{+2.0}_{-2.5}$ &27.0&  $12.3$ & $13.1$ &  0.9    \\

\hline
\end{tabular}
\caption{Parameters for the best-fit subhalo models, which assume a threshold halo concentration (a maximum for the unobscured sample, and a minimum for the obscured sample). 
}  
\label{table:ctable} 
\end{table}

\subsection{Clustering versus Black Hole Mass}

\begin{figure*}
\centering
\includegraphics[width=.49\textwidth]{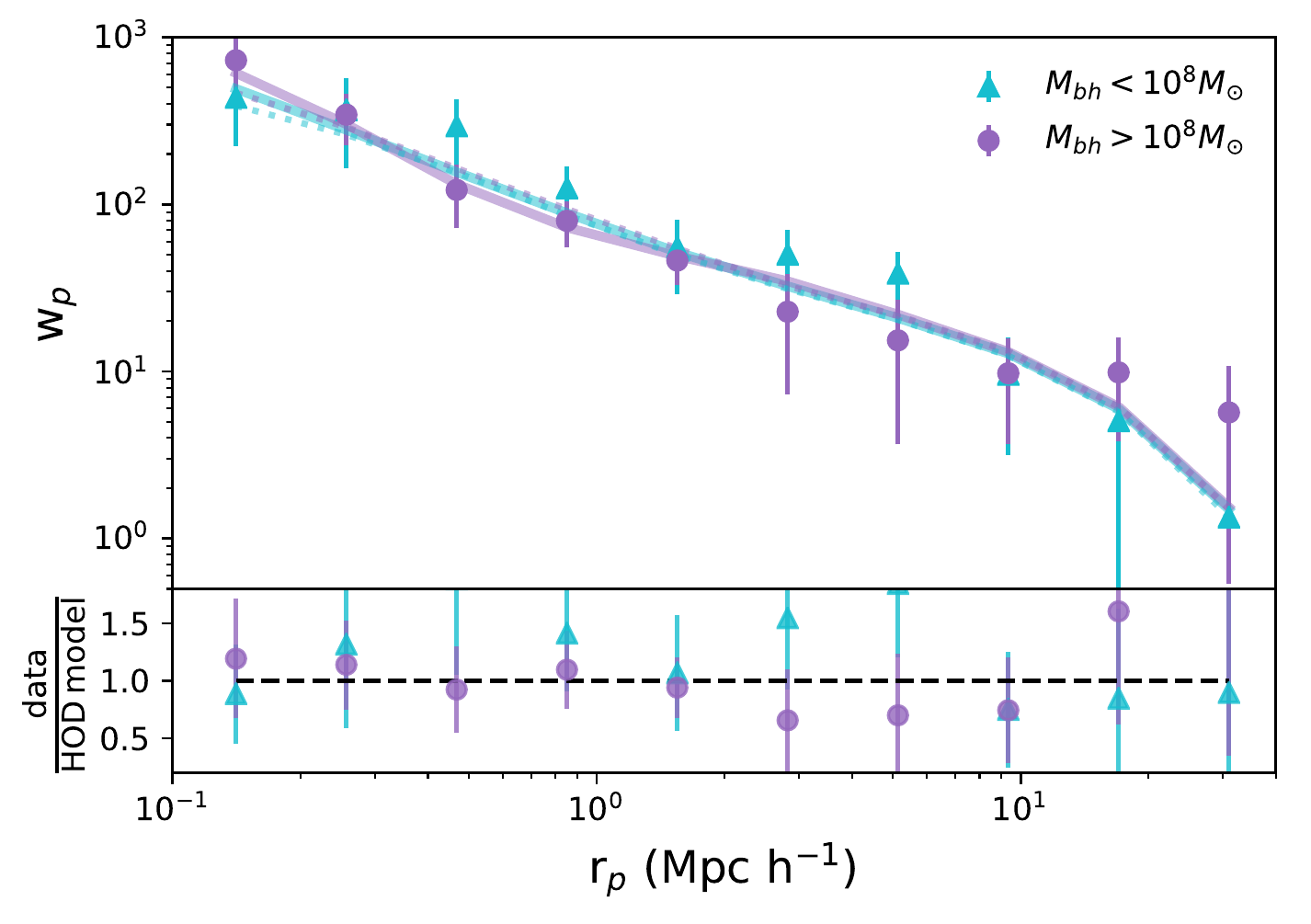}
\includegraphics[width=.49\textwidth]{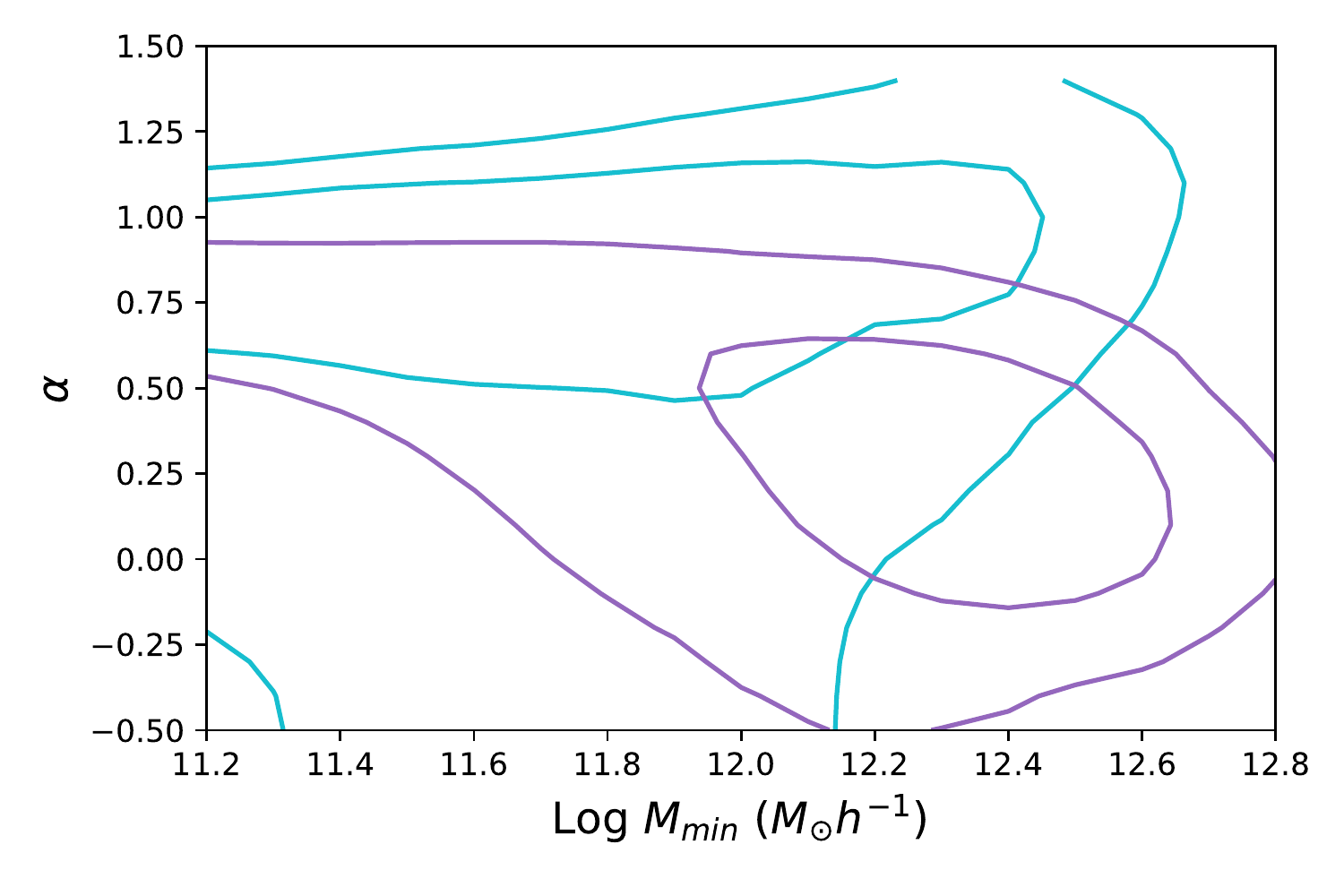}
\includegraphics[width=.75\textwidth]{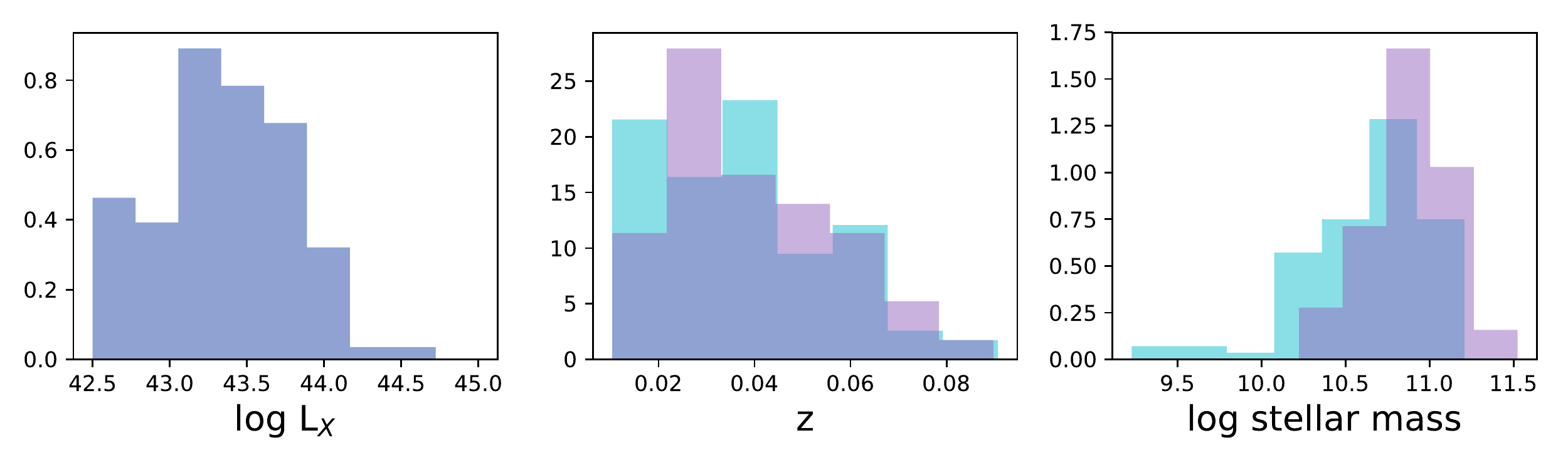}
\caption{Upper left: projected cross-correlation function in two bins of black hole mass, $M_{\rm bh}<10^{8} M_{\odot}$ (cyan) and $M_{bh}>10^{8} M_{\odot}$ (purple), with corresponding HOD model fits (solid lines) and subhalo models (dotted lines). Upper right: $\Delta \chi^{2}$ contour map of the HOD fit for each mass bin. Lower panels: distributions of the log of the X-ray luminosity (matched), redshift, and log of the host galaxy stellar mass.}
\label{fig:mbh}
\end{figure*}

Figure \ref{fig:mbh} shows the results of the correlation function measurements and HOD fitting for the AGN sample divided into two bins of black hole mass. We again randomly down-sampled each bin in order to avoid selection effects. The differences between the two samples are not significant; there is a $\sim 1\sigma$ difference in $\alpha$, in the sense that the satellite slope is shallower for large black hole masses than for the smaller ones, with best-fit values $0.2\pm0.5$ and $0.9\pm0.3$, respectively. 
A satellite power-law slope of 0 is consistent with the population residing purely in central galaxies; this is within the uncertainties, given such large black hole mass bin sizes and the limited sample size.
(While the $\chi^2_{\nu}$ is large for the small black hole bin (2.6), it should be noted that it becomes 1.3 with the same best-fit parameters after removing one data point.)

While the correlation between black hole mass and halo mass \citep[e.g.,][]{Silk:1998,El-Zant:2003,Booth:2010} would predict that the larger black hole bin would have a larger bias, as was found in \cite{Krumpe:2015}, we find no significant difference. The median halo masses for each bin are $\log~M_{h}/M_{\odot}h^{-1}=12.6\pm0.3$ and $12.8\pm0.3$, for small and large black holes, respectively.

Our results may suggest that larger black holes are less likely to reside in satellite galaxies, which would make sense assuming a correlation between the mass of the black hole and the mass of its host \emph{subhalo}. More data are needed to conclusively confirm this.

\section{Discussion}
\subsection{Environments of Local AGN}
We have cross-correlated hard X-ray selected AGN with 2MASS near-infrared-selected galaxies to constrain how an unbiased sample of local AGN occupy their halos. 
Analyzing the sample in terms of an HOD model, we find that the number of AGN hosted in a halo roughly scales with halo mass, as is the case for the overall galaxy population. 
This is inconsistent with the notion that AGN are predominantly in central galaxies \citep[e.g.,][]{Starikova:2011,Richardson:2013}, as our results suggest a significant fraction of AGN are in satellites. This agrees with several recent studies \citep{Allevato:2012,Oh:2014,Silverman:2014}. 
Additionally, using a subhalo-based model that corrects for catalog incompleteness, we find that the host galaxy stellar mass distribution can determine the environments of AGN on average, via the stellar mass-subhalo mass relation \citep{Behroozi:2010}. This was also found when comparing predictions of this model with the weak gravitational lensing signal of X-ray-selected AGN in COSMOS \citep{Leauthaud:2015}.

The typical halo mass found for the BASS AGN with our HOD analysis, $\log M_h/M_{\odot}~h^{-1} = 12.8$, lies between those typically found for soft X-ray-selected AGN ($\log M_h/M_{\odot}~h^{-1} \sim 13$) and optically selected AGN ($\log M_h/M_{\odot}~h^{-1} \sim 12.5$), and thus is broadly consistent with earlier results from AGN clustering studies across a large range of luminosity and redshift \citep{Croom:2005,Gilli:2005,Gilli:2009,Ross:2009,Shen:2009,Allevato:2011,Krumpe:2012,Allevato:2014}. The typical halo masses of BASS AGN correspond to galaxy group environments. 

\subsection{Obscured versus Unobscured Environments}
We split our sample in two bins of $N_{\rm H}$ to test whether AGN with different column densities (i.e., obscured versus unobscured) live in different environments, for samples matched in luminosity and redshift, in order to avoid bias in the observed volume; we note that the host galaxy stellar mass distributions are also similar (Figure \ref{fig:t1t2_edd}). We find differences in their correlation functions, predominantly on small scales. Our HOD fits suggest that obscured AGN live in more massive halos and in denser environments than unobscured AGN. 

The simplest unification models attribute obscuration to the circumnuclear material absorbing the radiation produced in the broad-line region. In that case, whether the AGN is observed as obscured or unobscured depends only on viewing angle \citep{Urry:1995}, which means the halo-scale environments should be the same (statistically) for both populations. Although we now know that circumnuclear geometry is not the only factor, as there is a dependence of luminosity and $\lambda_{\rm Edd}$ on covering factor \citep[e.g.,][]{Ricci:2017}, the analysis of our matched samples shows that these factors are not biasing our results.

Large column densities can also come from the host galaxy; for example, from a random molecular cloud that happens to lie along the line of sight or from the orientation of the galaxy disk. In the present case, only about 5\% of the sample lies in an edge-on host galaxy, and the results do not change when those AGN are removed. Another possibility is from inflowing gas following a merger \citep[e.g.,][]{Hopkins:2008,Kocevski:2015,Ricci:2017C}.
In general, because the probability of galaxy interactions depends on the environment \cite[e.g.,][]{Shen:2009,Jian:2012}, it is possible that either major mergers or smaller galaxy interactions play a role in causing the clustering difference. 

\cite{Dipompeo:2017B} found a similar clustering difference on large scales with \emph{WISE} infrared-selected AGN at $z \sim 1$. They interpreted their results as obscuration being an evolutionary phase of merger-driven quasar fueling, in which the quasar is first obscured, followed by an unobscured phase after gas `blow-out'. The resulting observations of obscured AGN living in larger halos would be a selection effect based on this model, by using luminosity-limited samples. Assuming this scenario for our AGN sample, the halo mass differences should be minimal at these low luminosities ($\sim 0.2$ dex of $M_{\odot}h^{-1}$)---much less than our results based on the HOD analysis. It is unlikely that major mergers trigger these low-luminosity, low-redshift AGN. Indeed, only 8\% of BAT AGN are in the final phases of major mergers, where obscuration is found to peak \citep{Koss:2010}.

The shallow satellite power-law slope of unobscured AGN, $\alpha$, obtained from the HOD analysis, would suggest that the fraction of galaxies that host unobscured AGN drops as a function of halo mass. This would mean that unobscured AGN avoid the richest clusters. Because high velocity encounters in the largest clusters disfavor galaxy mergers, perhaps a higher fraction of unobscured AGN were triggered by an earlier major merger, such that there was sufficient time to clear the surrounding gas and dust. Using the analytical function of the instantaneous galaxy merger rate from \cite{Shen:2009B}, we estimate that major mergers occur roughly four times more often in halos of the average mass hosting unobscured AGN than in halos hosting obscured AGN around $z\sim 0.1$. However, we compared the halo masses for obscured and unobscured AGN from the 2MASS group catalog \citep{Lu:2016} for the sources with counterparts in 2MASS, and found no evidence that obscured AGN live in preferentially larger halos.

Alternatively, we have shown that a difference in halo concentration, opposed to differing halo occupation distributions and/or typical halo masses, fits the data equally well. Highly concentrated halos of a given mass would have a high concentration of satellite galaxies, and therefore have a higher probability of galaxy interactions (i.e., minor mergers and encounters, as opposed to major mergers that predominantly occurred at high redshifts). Indeed, \cite{Jiang:2016} found that SDSS Type 2 satellites were more concentrated than Type 1 satellites, and \cite{Villarroel:2014} calculated an enhanced number of SDSS Type 2 vs. Type 1 companions around Type 2 AGN. The excess of Compton-thick BASS AGN in mergers would also support this scenario \citep{Koss:2016}. However, after removing clear cases of mergers and interactions in the obscured sample by visual inspection, the clustering differences slightly increased rather than decreased --- the opposite of what this scenario would predict. Additionally, the unobscured sample is the one that is more inconsistent with the clustering statistics expected for its stellar mass distribution, suggesting unobscured AGN have more environmental dependencies than obscured AGN. 

Instead, the observed difference between the clustering of obscured and unobscured AGN may be due to a difference in their host halo assembly histories. Halo concentration correlates with formation epoch, and so unobscured AGN tend to reside in halos that were assembled more recently in cosmic time than halos hosting obscured AGN. This means that the merging of their subhalos, and hence the merging of the galaxies within these subhalos, occurred around $z\sim 1$, opposed to at higher redshift for obscured AGN host halos. Therefore, the progenitors of $z=0$ unobscured AGN underwent major merger events statistically more recently than obscured AGN. If the major mergers triggered a powerful quasar that blew away much of the surrounding gas and dust, then it would explain the lower column densities we see in AGN in recently formed halos. Perhaps obscured AGN host halos had, on average, a more quiescent history dominated by secular processes, allowing nuclear obscuring material to remain. This scenario, with the different host halo histories rather than AGN triggering processes, explains the distinct clustering signatures we see for unobscured and obscured AGN at $z\sim0$.
However, it is uncertain if this explanation is consistent with higher redshift studies \citep[e.g.,][]{Allevato:2014}; an investigation of obscured vs. unobscured AGN clustering with samples of matched stellar mass distributions across a wide range of redshift (and luminosity) is needed.

\subsection{Possible Dependence of Environment on Black Hole Mass} 

There is a small ($\sim 1\sigma$) difference between the clustering of AGN with black holes of mass $<10^{8} M_{\odot}$ and $>10^{8} M_{\odot}$. 
The flatter satellite power-law slope indicated by our analysis may suggest that larger black holes tend to lie in central galaxies rather than satellites, while smaller black holes tend to lie in satellites. A correlation between the SMBH and the mass of its subhalo goes in the right direction. However, more data are clearly necessary to confirm this weak signal.

\section{Summary}

In this study, we characterized the environments of a sample of accreting SMBHs unbiased toward obscuration by measuring the cross-correlation function of BASS AGN and 2MASS galaxies. We compared our results to mock samples created from simulations in order to model how AGN occupy their host dark matter halos.

From fitting an HOD model to the cross-correlation function of the full sample, and by comparing with a subhalo model that assumed only stellar mass determines clustering statistics, we concluded that BASS AGN, on average, occupy dark matter halos consistently with the overall inactive galaxy population.

However, subsamples based on column density and black hole mass have differing clustering statistics. We found that absorbed AGN reside in denser environments than unabsorbed AGN, despite no significant difference in their luminosity, redshift, or stellar mass distributions. Our subhalo model analysis suggests they may reside in halos with statistically different concentrations/assembly histories. The alternative interpretation from the HOD analysis --- that they have systematically different halo occupation distributions and host halo masses --- seems to contradict the finding that stellar mass drives the clustering amplitude.
Lastly, we found a hint that 
a larger fraction of high-mass black holes ($M>10^{8} ~M_{\odot}$) reside in central galaxies than for lower mass black holes.

\acknowledgments
M.P. would like to thank Andrew Hearin for helpful discussions.
M.P., N.C., and C.M.U acknowledge support from NASA-SWIFT GI: Nr. 80NSSC18K0505, NSF grant 1715512, NASA CT Space Grant, and Yale University. M.K. acknowledges support from NASA through ADAP award NNH16CT03C, and C.R. acknowledges support from FONDECYT 1141218, CONICYT PAI77170080, Basal-CATA PFB--06/2007, and the China-CONICYT fund.

\software{CorrFunc \citep{Sinha:2017},
Halotools \citep{Hearin:2017}, 
Astropy \citep{Astropy:2013}, 
Matplotlib \citep{matplotlib:2007}.}

\bibliographystyle{yahapj}
\bibliography{references}

\end{document}